\documentclass[aps,showpacs,preprintnumbers,amsmath, amssymb]{revtex4}

\usepackage{float}
\usepackage{graphics,epsfig}
\usepackage{graphicx}
\usepackage{dcolumn}
\usepackage{bm}

\begin{document}
\baselineskip=0.8 cm

\title{{\bf Analytical investigations on formations of hairy neutral reflecting shells in the scalar-Gauss-Bonnet gravity}}
\author{Yan Peng$^{1}$\footnote{yanpengphy@163.com}}
\affiliation{\\$^{1}$ School of Mathematical Sciences, Qufu Normal University, Qufu, Shandong 273165, China}

\vspace*{0.2cm}
\begin{abstract}
\baselineskip=0.6 cm
\begin{center}
{\bf Abstract}
\end{center}

We study scalarization of spherically symmetric neutral reflecting shells
in the scalar-tensor gravity. We consider neutral static massless scalar fields non-minimally coupled
to the Gauss-Bonnet invariant. We obtain a relation representing the existence regime of
hairy neutral reflecting shells. For parameters unsatisfying this relation, the massless scalar
field cannot exist outside the neutral reflecting shell. In the parameter region where this relation
holds, we get analytical solutions of scalar field hairs outside neutral reflecting shells.

\end{abstract}

\maketitle
\newpage
\vspace*{0.2cm}

\section{Introduction}

One well known property of classical black holes is the
famous no hair theorem, which states that spherically symmetric black holes
cannot support static scalar field hairs in the asymptotically
flat background, see references \cite{Bekenstein}-\cite{sn4} and reviews \cite{Bekenstein-1,CAR}.
The belief in this no hair behavior is partly based on the existence
of black hole absorbing horizons.
According to some candidate quantum-gravity models,
quantum effects may prevent the formation of stable black-hole horizons \cite{UH1,UH2,UH3,UH4,UH5}.
And horizonless compact objects with reflective
boundary conditions have been proposed as alternatives to
the familiar (classical) black-hole spacetimes \cite{RS1,RS2,RS3,RS4,RS5,RS6,RS7}.
So it is interesting to study properties of horizonless
reflecting objects.

Interestingly, no hair theorem also holds in such horizonless
reflecting object backgrounds. Hod firstly proved that massive static scalar field hairs cannot
form in the background of neutral horizonless reflecting objects \cite{Hod-1}.
This no hair theorem for neutral horizonless reflecting objects was also
extended to the case of massless scalar field hairs \cite{Hod-5,Yan Peng-1}.
Considering a positive cosmological constant,
it was found that the no hair theorem still holds in the background
of neutral horizonless reflecting objects \cite{SBS}.
This no hair theorem for composed system of scalar fields
and neutral horizonless reflecting objects was further generalized
by including couplings between scalar fields and Ricci curvature \cite{Hod-5,Hod-6}.
However, when horizonless reflecting objects are charged,
analytical and numerical results showed that
scalar field hairs can exist \cite{Hod-2}-\cite{BKM}.
From front progress, we conclude that no static scalar field hair behavior is a very general property
in the background of neutral horizonless reflecting objects.

In other modified gravities, whether static scalar field hairs could
exist outside neutral horizonless reflecting objects
is a question to be answered.
On the other side of black holes, usual ways to introduce scalar hairs
are considering stationary scalar fields or
adding a confinement to the system \cite{vn1}-\cite{vn10}.
Recently, a novel approach to trigger black hole scalar hairs was provided
by considering non-minimal couplings between scalar fields
and the Gauss-Bonnet invariant \cite{SGB1,SGB2,SGB3,SGB4,SGB5,SGB6,SGB7}.
Moreover, it was found that this scalar-Gauss-Bonnet coupling
can lead to scalar condensations in various black hole models
\cite{SGB8,SGB9,SGB10,SGB11,SGB12,SGB13,SGB14,SGB15,SGB16}.
Inspired by these black hole properties, in the background of
neutral reflecting compact stars, we have constructed
scalar hairy configurations by including scalar-Gauss-Bonnet
couplings with numerical methods \cite{YPSC}.
In particular, reflecting shell backgrounds usually allow
fully analytical studies, which showed that neutral reflecting
shells cannot support static scalar hairs \cite{Hod-2,Hod-3}.
As a further step, it is interesting to examine whether scalar fields can
condense outside neutral reflecting shells in the
model generalized by including scalar-Gauss-Bonnet couplings.

This work is organized as follows.
We firstly construct a system with static massless scalar fields
outside neutral horizonless reflecting shells
in the scalar-Gauss-Bonnet gravity.
Then we obtain a relation representing existence regime
of hairy shells. For parameters satisfying this relation,
we get analytical solutions of scalar field hairs outside
neutral reflecting shells. The analytical solutions
presented in this paper are valid only in the
linearized regime of the scalar fields.
At last, we give the main conclusion.

\section{Scalar condensation behaviors around neutral Dirichlet reflecting shells}

\subsection{A characteristic relation for scalar hairy neutral reflecting shells}

We now write down the model with static massless scalar
fields non-minimally coupled to the Gauss-Bonnet invariant. The
Lagrangian density of this scalar-tensor gravity is described by
\cite{SGB1,SGB2,SGB3,SGB4,SGB5,SGB6,SGB7}
\begin{eqnarray}\label{lagrange-1}
\mathcal{L}=R-|\nabla_{\mu} \psi|^{2}+f(\psi)\mathcal{R}_{GB}^{2},
\end{eqnarray}
where R is the Ricci curvature,
$\psi(r)$ is a real scalar field, $f(\psi)$ is the coupling function
and $\mathcal{R}_{GB}^{2}$ is the source term.
In the linear regime, without generality, we can take
the coupling function in the form
\begin{eqnarray}\label{lagrange-1}
f(\psi)=\eta\psi^2
\end{eqnarray}
with $\eta$ describing the coupling strength \cite{SGB3,SGB4}.
The source term is the Gauss-Bonnet invariant given by
\begin{eqnarray}\label{lagrange-1}
\mathcal{R}_{GB}^{2}=R_{\mu\nu\rho\sigma}R^{\mu\nu\rho\sigma}-4R_{\mu\nu}R^{\mu\nu}+R^2.
\end{eqnarray}
When neglecting matter fields' backreaction on the metric,
the Gauss-Bonnet invariant term is
\begin{eqnarray}\label{lagrange-1}
\mathcal{R}_{GB}^{2}=\frac{48M^2}{r^6}.
\end{eqnarray}

We consider spherically symmetric static neutral spacetimes.
The metric ansatz in Schwarzschild
coordinates is of the form \cite{SGB4}
\begin{eqnarray}\label{AdSBH}
ds^{2}&=&-g(r)dt^{2}+\frac{dr^{2}}{g(r)}+r^{2}(d\theta^{2}+sin^{2}\theta d\phi^{2}),
\end{eqnarray}
where the metric function is
$g(r)=1-\frac{2M}{r}$ with $M$ corresponding to the ADM mass.
The shell radius is imposed at the radial coordinate $r=r_{s}$.
Since we concentrate on the horizonless spacetime,
the shell radii satisfy the relation $r_{s}>2M$.
The spherically symmetric angular coordinates are
labeled as $\theta$ and $\phi$.

With variation methods, we get the exact linearized scalar equation \cite{SGB1,SGB2,SGB3,SGB4,SGB5,SGB6,SGB7}
\begin{equation}
\nabla^{\nu}\nabla_{\nu}\psi+\eta \mathcal{R}_{GB}^{2} \psi =0.
\end{equation}

By employing the line element (5), the scalar equation
takes the form \cite{YPSC}
\begin{equation}
\psi''+(\frac{2}{r}+\frac{g'}{g})\psi'+\frac{\eta\mathcal{R}_{GB}^{2}}{g}\psi=0
\end{equation}
with $g=1-\frac{2M}{r}$ and $\mathcal{R}_{GB}^{2}=\frac{48M^2}{r^6}$.

In the limit case of $M\ll r_{s}$, the functions are $g(r)=1-\frac{2M}{r}\rightarrow 1$
and $g'(r)=\frac{2M}{r^2}\rightarrow 0$ as assumed in [29,30].
In the large-r regime, $g'(r)$ is neglected and $g(r)$ is set to be 1.
The presence of a coupling parameter $\eta$ is crucial for the
existence of a non-trivial analytical solution.
With the nonzero term $\eta\mathcal{R}_{GB}^{2}=\frac{48\eta M^2}{r^6}$, $\eta$ appears in the
scalar field equation. It means that we study the scalar condensation
in the large $\eta$ regime.
In this shell background, the equation (7) can be expressed as
\begin{equation}\label{BHg}
\psi''+\frac{2}{r}\psi'+\frac{48\eta M^2}{r^6}\psi=0.
\end{equation}

In order to solve the equation, we need boundary
conditions of the scalar field.
The asymptotic behavior of the massless scalar field
near the infinity boundary is
\begin{eqnarray}\label{InfBH}
&&\psi\propto\frac{1}{r}~~~for~~~r\rightarrow\infty.
\end{eqnarray}
So the infinity boundary condition is
\begin{equation}
\psi(\infty)=0.
\end{equation}

At the shell radius, we impose Dirichlet reflecting boundary conditions
that the scalar field vanishes. So the scalar field condition at the surface is
\begin{eqnarray}\label{InfBH}
\psi(r_{s})=0.
\end{eqnarray}

We introduce a new radial function $\tilde{\psi}=\sqrt{r}\psi$.
According to (8), $\tilde{\psi}$ satisfies the differential equation
\begin{eqnarray}\label{BHg}
r^2\tilde{\psi}''+r\tilde{\psi}'+(-\frac{1}{4}+\frac{48\eta M^{2}}{r^4})\tilde{\psi}=0.
\end{eqnarray}

With relations (9) and (11), we get boundary conditions
\begin{eqnarray}\label{InfBH}
&&\tilde{\psi}(r_{s})=0,~~~~~~~~~\tilde{\psi}(\infty)=0.
\end{eqnarray}

From boundary conditions (13), one deduces that
the function $\tilde{\psi}$ must possess one extremum point $r=r_{peak}$
in the range $(r_{s},\infty)$. At this extremum point, the scalar field satisfies relations \cite{Hod-1}
\begin{eqnarray}\label{InfBH}
\{ \tilde{\psi}'=0~~~~and~~~~\tilde{\psi} \tilde{\psi}''\leqslant0\}~~~~for~~~~r=r_{peak}.
\end{eqnarray}

Relations (12) and (14) yield the following inequality
\begin{eqnarray}\label{BHg}
-\frac{1}{4}+\frac{48\eta M^{2}}{r^4}\geqslant0~~~for~~~r=r_{peak}.
\end{eqnarray}

This inequality can be transformed into
\begin{eqnarray}\label{BHg}
\frac{\sqrt{\eta}M}{r^2}\geqslant \frac{1}{8\sqrt{3}}~~~for~~~r=r_{peak}.
\end{eqnarray}

Considering $r_{s}\leqslant r_{peak}$,
we conclude that scalar hairy shells should satisfy the relation
\begin{eqnarray}\label{BHg}
\frac{\sqrt{\eta}M}{r_{s}^2}\geqslant \frac{1}{8\sqrt{3}}.
\end{eqnarray}

\subsection{Construction of massless scalar field hairy neutral Dirichlet reflecting shells}

In this section, we apply analytical methods to get
solutions of scalar field hairs outside
Dirichlet reflecting shells in the scalar-Gauss-Bonnet gravity.
The general solutions of equation (12) can be
expressed with Bessel functions in the form \cite{Abramowitz}
\begin{eqnarray}\label{BHg}
\tilde{\psi}(r)=A\cdot
  J_{-\frac{1}{4}}(\frac{2\sqrt{3\eta}M}{r^2})+B\cdot
  J_{\frac{1}{4}}(\frac{2\sqrt{3\eta}M}{r^2})
\end{eqnarray}
with $A$ and $B$ as integral constants.

At the infinity, the solution (18) asymptotically behaves as
\begin{eqnarray}\label{BHg}
\tilde{\psi}(r)\propto A\cdot \sqrt{r}+B\cdot \frac{1}{\sqrt{r}}.
\end{eqnarray}
According to the condition (13), the first coefficient $A$
is zero: $A=0$. So the bound-state neutral massless scalar fields are
\begin{eqnarray}\label{BHg}
\psi=\sqrt{\frac{1}{r}}\tilde{\psi}(r)=
B\cdot \sqrt{\frac{1}{r}} J_{\frac{1}{4}}(\frac{2\sqrt{3\eta}M}{r^2}).
\end{eqnarray}

With the scalar reflecting condition (11), we get the characteristic scalar field equation
\begin{eqnarray}\label{BHg}
J_{\frac{1}{4}}(\frac{2\sqrt{3\eta}M}{r_{s}^2})=0.
\end{eqnarray}
If we find parameters satisfying (21), scalar field hairs exist.
Defining a new parameter $x=\frac{\sqrt{\eta}M}{r_{s}^2}$,
there is $x\geqslant \frac{1}{8\sqrt{3}}$ according to (17).
The remaining question is to solve the equation
\begin{eqnarray}\label{BHg}
J_{\frac{1}{4}}(2\sqrt{3}x)=0
\end{eqnarray}
in the region $x\geqslant \frac{1}{8\sqrt{3}}$.
With numerical methods, the condition (22) determines discrete values of $x_{i}$
\begin{eqnarray}\label{BHg}
\cdots~>x_{3}~>~x_{2}~>~x_{1}~=x_{min}~\geqslant~ \frac{1}{8\sqrt{3}}.
\end{eqnarray}
Fixing shell radii at $r_{si}=\frac{\eta^{\frac{1}{4}} M^{\frac{1}{2}}}{x_{i}^{\frac{1}{2}}}$,
the corresponding scalar field is $\psi\propto\sqrt{\frac{1}{r}}J_{\frac{1}{4}}(\frac{2\sqrt{3\eta}M}{r^2})$
in the form of (20). In Table I, we show various values of $r_{si}$ with respect to $i$.
We plot the first two solutions of scalar fields in the background of Dirichlet reflecting shells in Fig. 1.
The scalar fields start from zero at the shell radii and
asymptotically approach zero at the infinity.
\renewcommand\arraystretch{1.7}
\begin{table} [h]
\centering
\caption{Radii of Dirichlet reflecting scalar hairy shells}
\label{address}
\begin{tabular}{|>{}c|>{}c|>{}c|>{}c|>{}c|>{}c|}
\hline
$~~~~i~~~~$ & ~~~~1~~~~ & ~~~~2~~~~& ~~~~3~~~~& ~~~~4~~~~& ~~~~5~~~~\\
\hline
$~~~~r_{si}~~~~$ & ~~~~1.1161$\eta^{\frac{1}{4}}M^{\frac{1}{2}}$~~~~ & ~~~~0.7658$\eta^{\frac{1}{4}}M^{\frac{1}{2}}$~~~~& ~~~~0.6189$\eta^{\frac{1}{4}}M^{\frac{1}{2}}$~~~~& ~~~~0.5332$\eta^{\frac{1}{4}}M^{\frac{1}{2}}$~~~~& ~~~~0.4755$\eta^{\frac{1}{4}}M^{\frac{1}{2}}$~~~~\\
\hline
\end{tabular}
\end{table}

\begin{figure}[h]
\includegraphics[width=180pt]{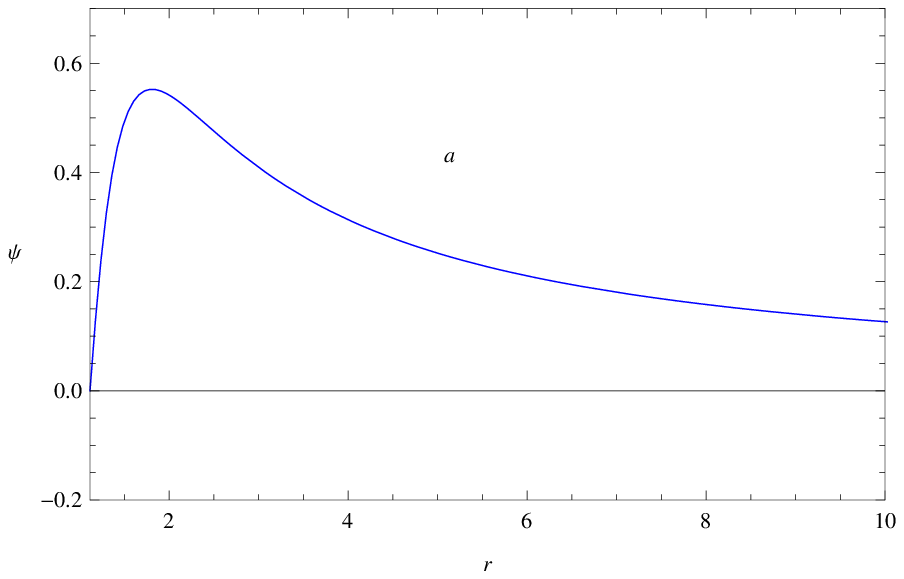}\
\includegraphics[width=180pt]{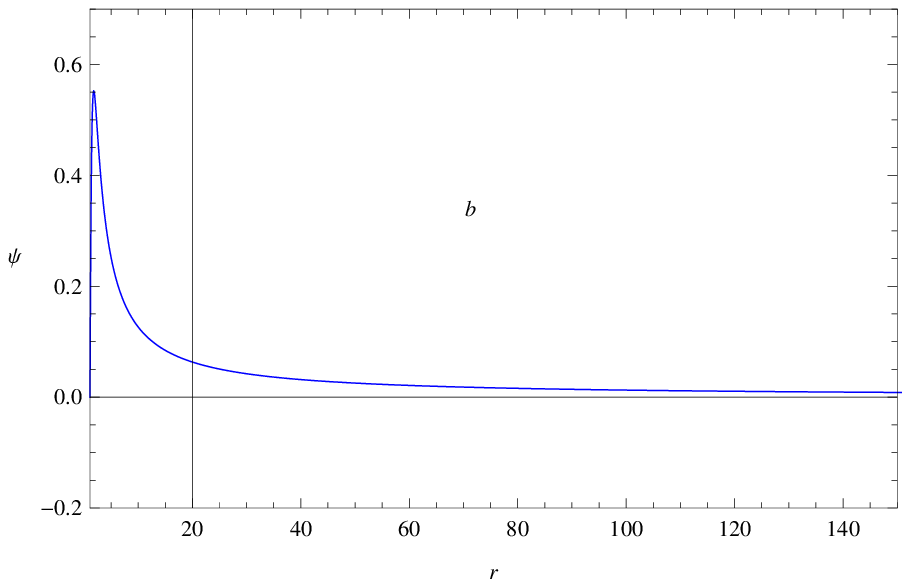}\
\includegraphics[width=180pt]{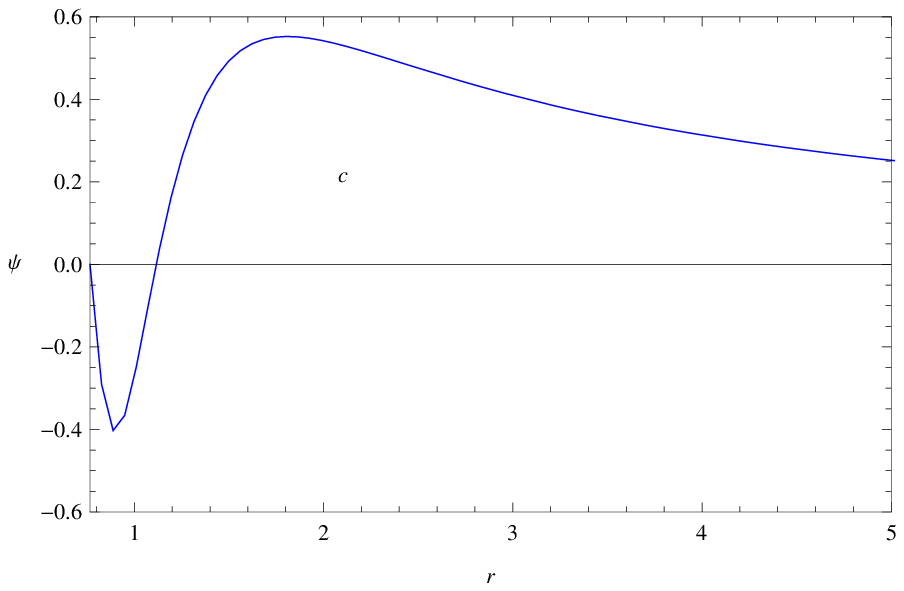}\
\includegraphics[width=180pt]{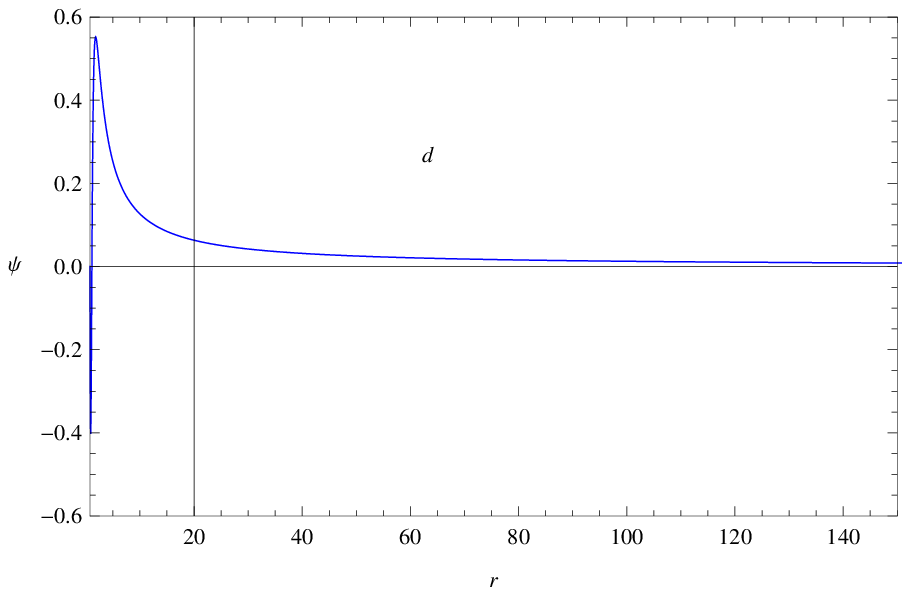}\
\caption{\label{EEntropySoliton} (Color online) We plot the scalar field solution
around Dirichlet reflecting shells. We take the value $\eta M^2=1$ with various $r_{s}$ as:
(a) the case $r_{s1}=1.1161$ in small region, (b) the case $r_{s1}=1.1161$ in large region,
(c) the case $r_{s2}=0.7658$ in small region, (d) the case $r_{s2}=0.7658$ in large region.}
\end{figure}

\section{Scalar condensation behaviors around neutral Neumann reflecting shells}

Now we turn to study scalar hair formations in the background
of neutral reflecting shells with Neumann surface boundary conditions.
At the surface, we impose the Neumann reflecting condition $\psi'(r_{s})=0$.
The derivative of the function $\tilde{\psi}$ satisfies boundary conditions
\begin{eqnarray}\label{BHg}
\widetilde{\psi}'(r_{s})=(\sqrt{r}\psi)'|_{r=r_{s}}=\frac{1}{2\sqrt{r_{s}}}\psi(r_{s})+\sqrt{r_{s}}\psi'(r_{s})
=\frac{1}{2\sqrt{r_{s}}}\psi(r_{s})=\frac{1}{2r_{s}}\sqrt{r_{s}}\psi(r_{s})=\frac{1}{2r_{s}}\widetilde{\psi}(r_{s}).
\end{eqnarray}
The case of $\widetilde{\psi}(r_{s})=0$ is just the model studied in section II. In this part,
we focus on the case of $\widetilde{\psi}(r_{s})\neq 0$.
In the case of $\widetilde{\psi}(r_{s})>0$,
the function $\tilde{\psi}$ increases to be more positive
around the surface and then decreases asymptotically to be zero.
In another case of $\widetilde{\psi}(r_{s})<0$,
the function decreases to be more negative around
the surface and then increases to be zero at the infinity.
For both cases, one extremum point $r=r_{peak}$ satisfying (14)
exists. Following analysis in part A of Section II,
for scalar hairy Neumann reflecting shells, we can easily
get the same relation as (17) in the form
\begin{eqnarray}\label{BHg}
\frac{\sqrt{\eta}M}{r_{s}^2}\geqslant \frac{1}{8\sqrt{3}}.
\end{eqnarray}

With the scalar field solution (20), we can express the Neumann reflecting condition as
\begin{eqnarray}\label{BHg}
\frac{d\psi}{dr}\mid_{r=r_{s}}=\frac{d}{dr}[\sqrt{r}\widetilde{\psi}]\mid_{r=r_{s}}
=\frac{d}{dr}[\sqrt{r}J_{\frac{1}{4}}(\frac{2\sqrt{3\eta}M}{r^2})]\mid_{r=r_{s}}=0.
\end{eqnarray}
The equation (26) can be solved through numerical methods.
In the parameter regime obeying (25), we obtain discrete values of shell radii
which can support the existence of static neutral massless scalar fields.
We give the discrete shell radii in Table II.
We also plot the first two solutions of scalar fields
outside Neumann reflecting shells in Fig. 2.
The scalar fields start with $\psi'(r_{s})=0$ at the
radii and asymptotically approaches zero in the large r region.
\renewcommand\arraystretch{1.7}
\begin{table} [h]
\centering
\caption{Radii of Dirichlet reflecting scalar hairy shells}
\label{address}
\begin{tabular}{|>{}c|>{}c|>{}c|>{}c|>{}c|>{}c|}
\hline
$~~~~i~~~~$ & ~~~~1~~~~ & ~~~~2~~~~& ~~~~3~~~~& ~~~~4~~~~& ~~~~5~~~~\\
\hline
$~~~~r_{si}~~~~$ & ~~~~1.8090$\eta^{\frac{1}{4}}M^{\frac{1}{2}}$~~~~ & ~~~~0.8992$\eta^{\frac{1}{4}}M^{\frac{1}{2}}$~~~~& ~~~~0.6823$\eta^{\frac{1}{4}}M^{\frac{1}{2}}$~~~~& ~~~~0.5720$\eta^{\frac{1}{4}}M^{\frac{1}{2}}$~~~~& ~~~~0.5022$\eta^{\frac{1}{4}}M^{\frac{1}{2}}$~~~~\\
\hline
\end{tabular}
\end{table}

\begin{figure}[h]
\includegraphics[width=180pt]{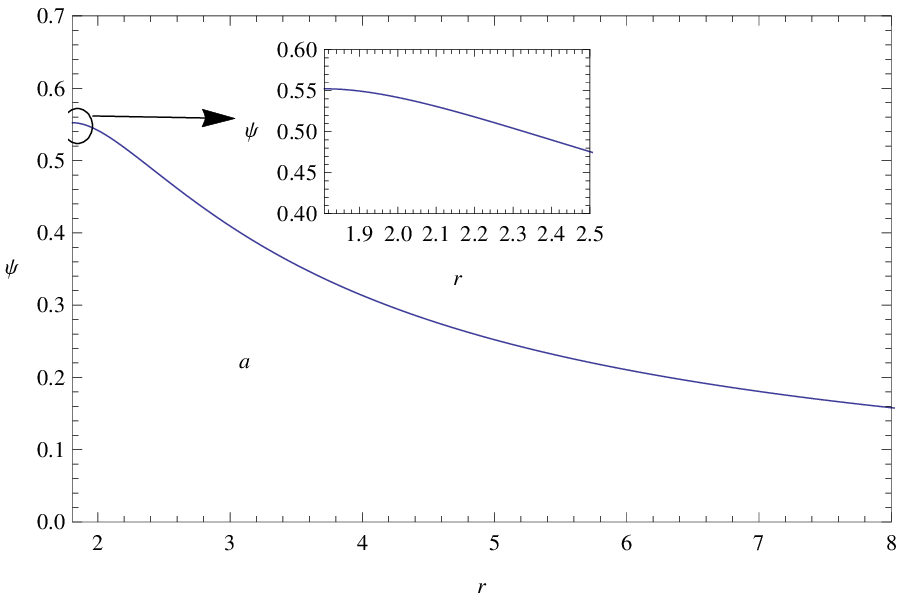}\
\includegraphics[width=180pt]{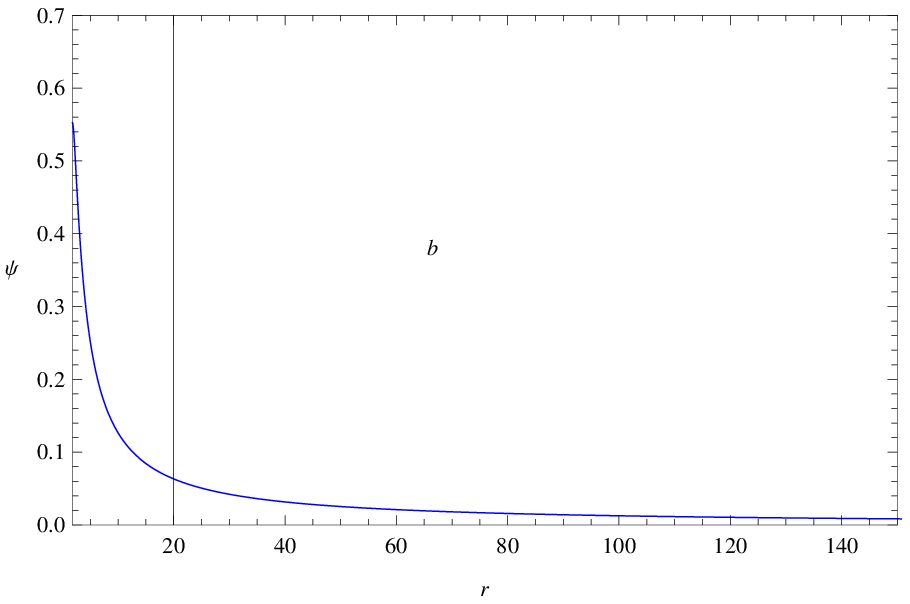}\
\includegraphics[width=180pt]{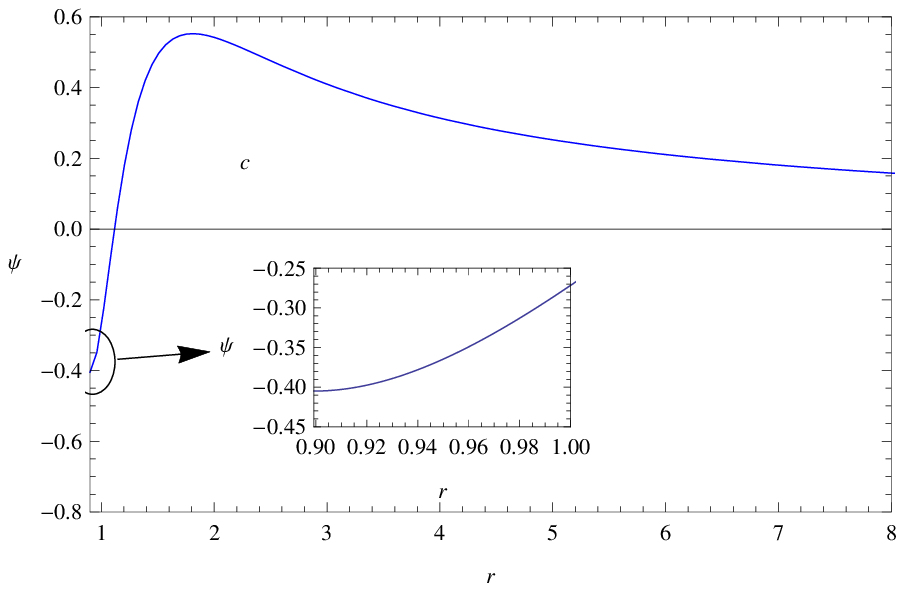}\
\includegraphics[width=180pt]{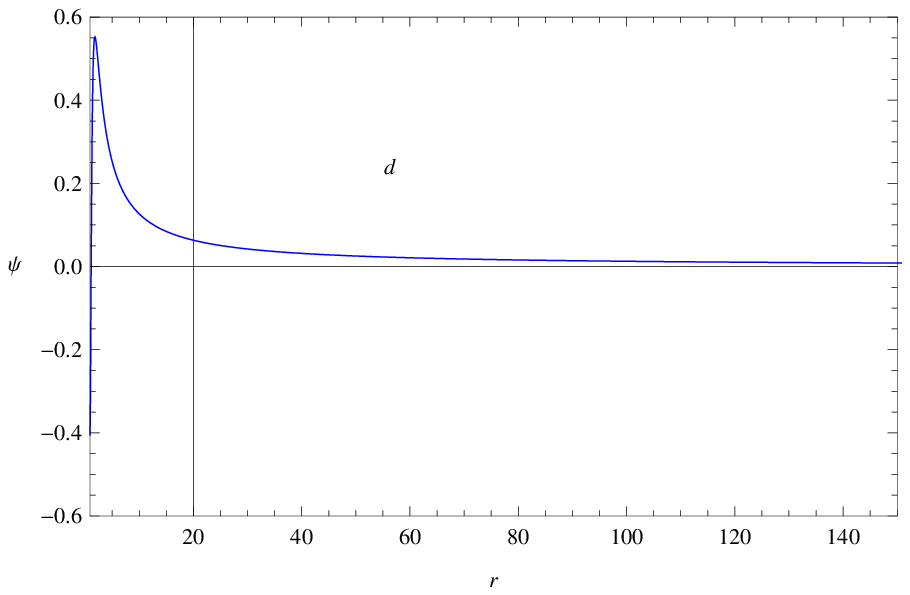}\
\caption{\label{EEntropySoliton} (Color online) We show behaviors of the scalar field
around Neumann reflecting shells. We take the value $\eta M^2=1$ with various $r_{s}$ as:
(a) the case $r_{s1}=1.8090$ in small region, (b) the case $r_{s1}=1.8090$ in large region,
(c) the case $r_{s2}=0.8992$ in small region, (d) the case $r_{s2}=0.8992$ in large region.}
\end{figure}

\section{Conclusions}

We studied condensations of static massless scalar fields
non-minimally coupled to the Gauss-Bonnet invariant
outside neutral reflecting shells.
At the shell radii, we imposed scalar reflecting boundary conditions.
We took two types of reflecting conditions,
which are Dirichlet and Neumann reflecting boundary conditions.
For both types of conditions, we analytically
obtained a characteristic relation for hairy shells in the form
$\frac{\sqrt{\eta}M}{r_{s}^2}\geqslant \frac{1}{8\sqrt{3}}$,
where $r_{s}$ is the shell radius, M is the shell mass and $\eta$ is the coupling parameter.
For parameters unsatisfying this relation, there is no scalar hair theorem.
For parameters obeying this relation, we obtained analytical
solutions of massless neutral scalar field hairs.

\begin{acknowledgments}

We would like to thank the anonymous referee for the constructive suggestions to improve the manuscript.
This work was supported by the Shandong Provincial Natural Science
Foundation of China under Grant No. ZR2018QA008. This work was also
supported by a grant from Qufu Normal University of China under
Grant No. xkjjc201906.

\end{acknowledgments}

\end{document}